\documentclass[a4paper,11pt]{article}
\usepackage{pos}

\newcommand{\ket}[1]{ {#1} \rangle}
\newcommand{\bq}{\mbox{\boldmath $q$}}
\newcommand{\bk}{\mbox{\boldmath $k$}}
\newcommand{\bp}{\mbox{\boldmath $p$}}
\newcommand{\Tr}{{\mbox{\rm Tr}}}
\title{Production of $\eta_{c}(1S,2S)$ in $e^+ e^-$ and $pp$ collisions}

\author*[a]{Izabela Babiarz}
\author[a]{Wolfgang Sch{\"a}fer}
\author[a]{Antoni Szczurek}

\affiliation[a]{ Institute of
Nuclear Physics Polish Academy of Sciences,\\ PL-31342 Krakow}

\emailAdd{Izabela.Babiarz@ifj.edu.pl}
\emailAdd{Wolfgang.Schafer@ifj.edu.pl}
\emailAdd{Antoni.Szczurek@ifj.edu.pl}
\begin{document}
\abstract{
We derive the light-front wave function (LFWF)
representation of the $\gamma^{\star} \gamma^{\star} \to \eta_{c} (1S),\eta_{c}(2S)$
transition form factor $F(Q^2_1,Q^2_2)$ for two virtual photons
in the initial state. For the LFWF, we use different models
obtained from the solution of the Schr\"odinger equation for
a variety of $c\bar{c}$ potentials. We compare our results to
the BaBar experimental data for the $\eta_{c}(1S)$
transition form factor, for one real and one virtual photon.
We observe that the onset of the asymptotic behaviour is strongly delayed
and discuss applicability of the collinear and/or massless limit.

In addition, we present a thorough analysis of $\eta_{c}(1S,2S)$
quarkonia hadroproduction in $k_{\perp}$-factorisation in the framework of
the light-front potential approach for the quarkonium wave function.
The off-shell matrix elements for the $g^{\star} g^{\star} \to \eta_{c} (1S,2S)$ vertices
are derived. We discuss the importance of taking into account the gluon
virtualities.
We present the transverse momentum distributions of $\eta_c$ for
several models of the unintegrated gluon distributions.
Our calculations are performed for four distinct parameterisations
for the $c\bar{c}$ interaction potential consistent with the meson spectra.
We compare our results for $\eta_{c}(1S)$ to measurements by the
LHCb collaboration and present predictions for $\eta_{c}(2S)$ production.
}

\FullConference{%
  40th International Conference on High Energy physics - ICHEP2020\\
  July 28 - August 6, 2020\\
  Prague, Czech Republic (virtual meeting)
}


\maketitle

\section{Introduction: Description of the mechanism $\gamma^{*} \gamma^{*} \rightarrow \eta_{c}(1S,2S)$}
\label{sect:Intro}

\begin{figure}[h!]
\centering \includegraphics[width= 0.4\linewidth]{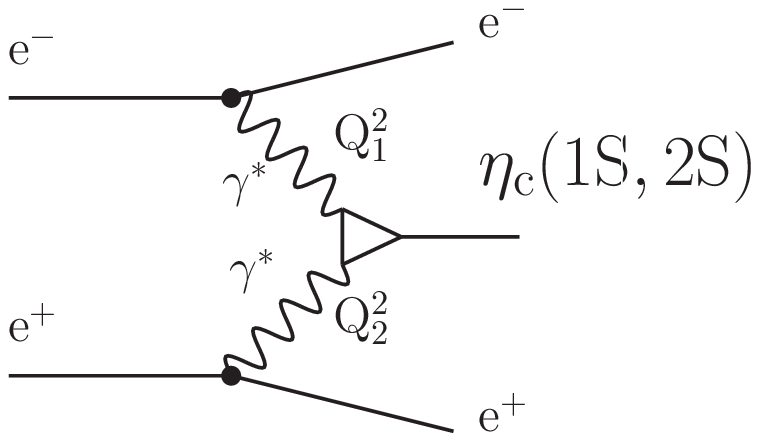} \quad
\centering \includegraphics[width = 0.4\linewidth]{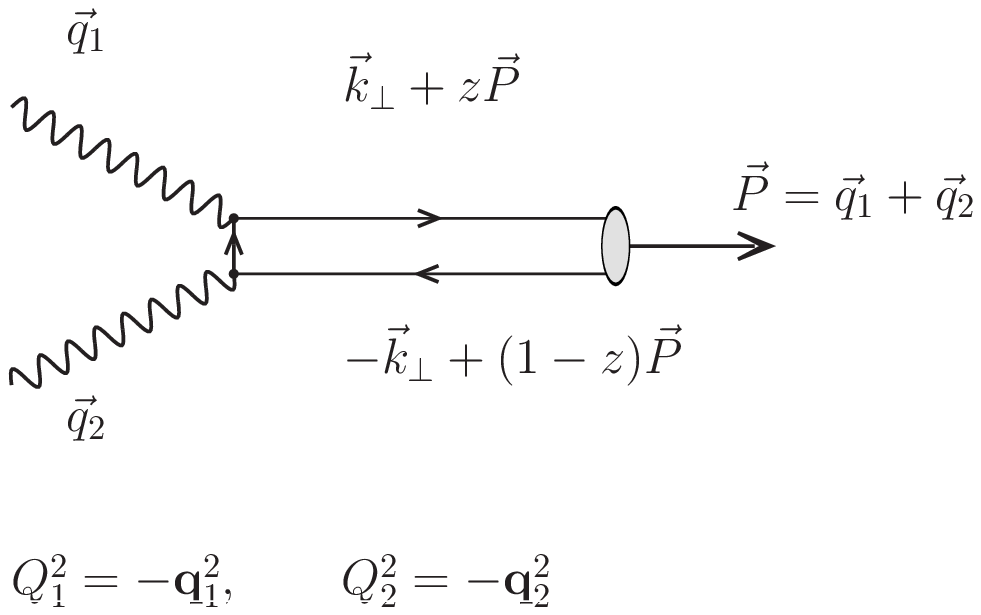}
\caption{Generic diagram for production of $\eta_c (1S)$ or the first
  excited state $\eta_c (2S)$. On the right hand side the 
the momenta involved are shown explicitly.
}
\label{fig:generic_diag}
\end{figure}

Complementary information for meson structure in quantum chromodynamics can be provided by
the study of electromagnetic form factors as well as meson-photon transition
form factors. Over a span of several years the attention has been paid mostly 
on the case of light pseudoscalar meson-photon transition form factors e.g. $\eta$,$\eta'$, $\pi^0$. 
Similar analysis have been performed for $\eta_c$ production. The mass
of the $\eta_c$ assure a hard scale and validate the perturbative
approach even for zero virtuality. Here we will focus on calculating
transition form factor for both virtual photon in the light-front
frame. For illustration in Fig.~\ref{fig:generic_diag} we present the
generic diagram for the process and  
the particle momenta involved into corresponding calculation.

The general form of the photon-photon fusion amplitude reads:

\begin{eqnarray}
{\cal M}_{\mu \nu}(\gamma^{*}(q_1) \gamma^{*}(q_2) \to \eta_c) 
= 4 \pi \alpha_{\rm em} \, (-i) \varepsilon_{\mu \nu \alpha \beta} q_1^\alpha q_2^\beta \, F(Q_1^2, Q_2^2) \,,
\end{eqnarray}

where $Q_1^2, Q_2^2$ are photon virtualities and light-front representation of the transition form factor:
\begin{eqnarray}
F(Q_{1}^{2}, Q_{2}^{2}) &=& e_{c}^{2} \sqrt{N_{c}} 4 m_{c}
 \cdot \int{\frac{dz d^2 \textbf{k}}{z(1-z) 16 \pi^3} \psi(z,\textbf{k})}\\
 & &{\Big\{} \frac{ 1-z}{(\textbf{k}- (1-z) \textbf{q}_2)^{2} + z (1-z) \textbf{q}_{1}^{2} + m_{c}^{2}}
+ \frac{z}{(\textbf{k} + z \textbf{q}_2 )^2 + z (1-z) \textbf{q}_1^2 + m_c^2}\nonumber
{\Big \}}. \,
\end{eqnarray}

\begin{figure}
\centering\includegraphics[width=0.39\linewidth]{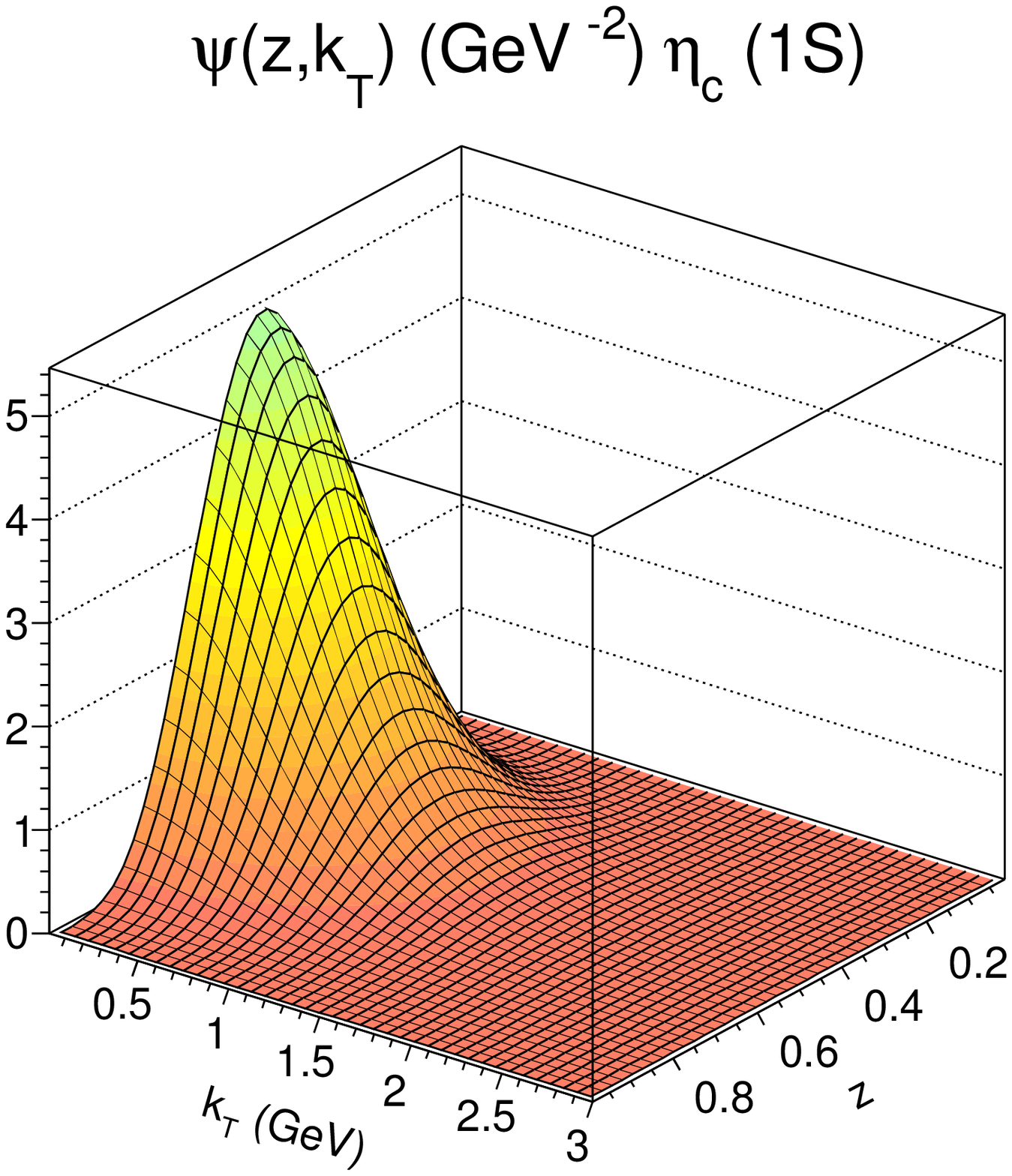}
\centering\includegraphics[width=0.39\linewidth]{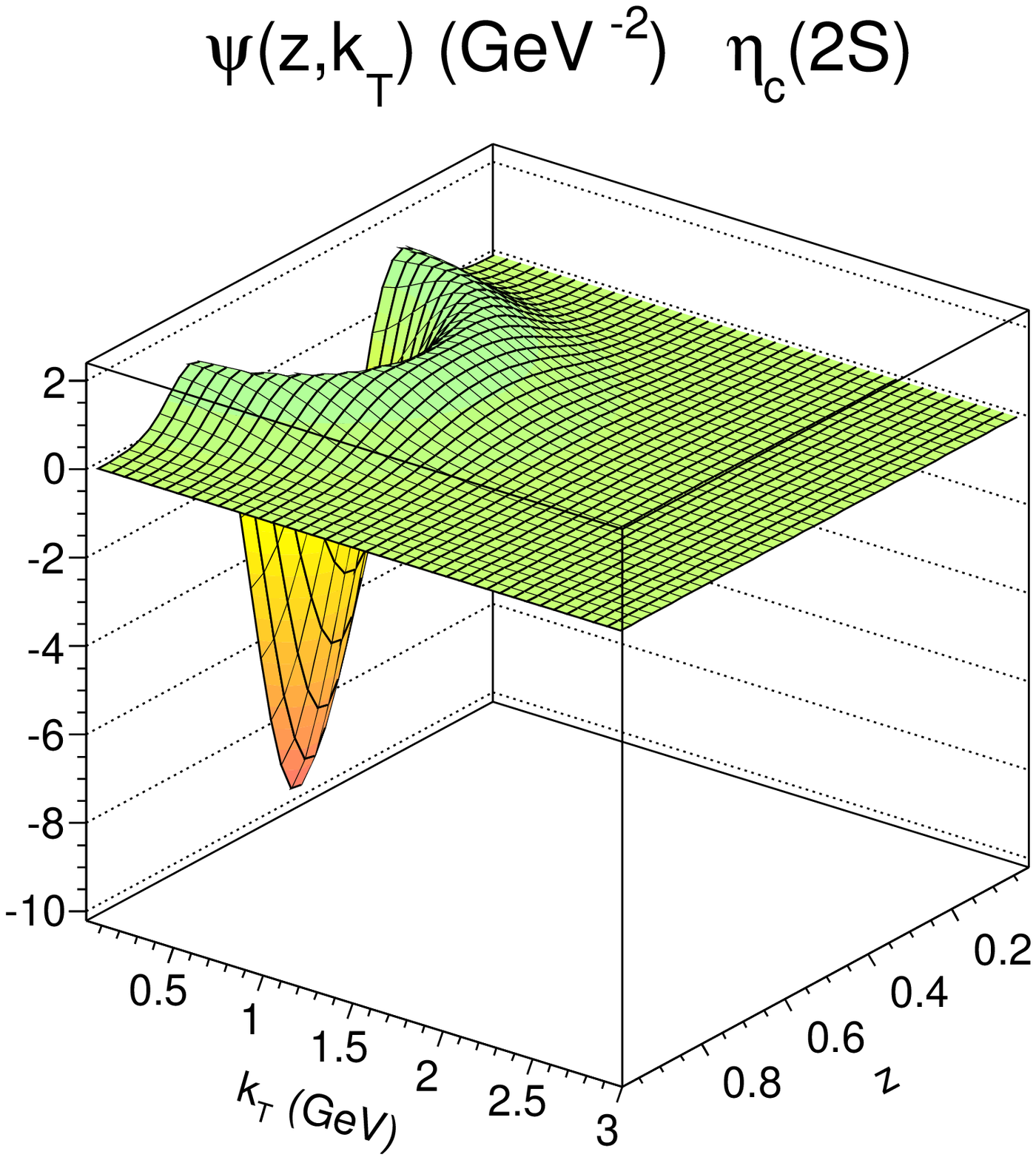}
\caption{Radial light-front wave function obtained for the 
Buchm\"uller-Tye potential.}
\label{fig:but_psi}
\end{figure}

In order to construct the form factor $F(Q_{1}^{2}, Q_{2}^{2})$, we have used
light-front wave functions $\psi(z,\bk)$ (see Fig.~\ref{fig:but_psi}).
These wave functions are obtained in a few steps, the first step is to solve 
the Schr\"odinger equation for several model of $c\bar{c}$ potential
and then the obtained non-relativistic radial space wave function $u(r)$ are transformed to momentum space $u(p)$ (for more details see \cite{Babiarz_FF, Cepila}). Due to Terentev prescription: $\textbf{p} = \textbf{k}, \, p_z = (z - {1/2}) M_{c \bar c}$  valid for weakly bound system we can rewrite:

\begin{eqnarray}
\Psi_{\lambda\bar{\lambda}}(z,\textbf{k}) = \bar{\textsc{u}}_{\lambda}(zP_{+},\textbf{k})\gamma_{5}
\textsc{v}_{\bar{\lambda}}((1-z)P_{+},-\textbf{k})\,\psi(z,\textbf{k})\, , \quad 
\psi(z,\textbf{k}) = {\pi \over {\sqrt{2 M_{c \bar c}}}} { u(p) \over p} \, .
\end{eqnarray}

Here $M_{c \bar c} = \sqrt{(\textbf{k}^2 + m_c^2)/(z(1-z))}$ is the invariant mass of the $c \bar c$ state which depends
on momenta of quarks. 
It is worth to notice that we treat $\eta_c$ meson as a bound state of $c\bar{c}$ and therefore we assume that dominant component in the Fock-state expansion comes from  $c\bar{c}$:

\begin{eqnarray}
|\ket{\eta_c; P_+, \mathbf{P}}= \sum_{i,j,\lambda, \bar \lambda}
{\delta^i_j \over \sqrt{N_c}} \, 
\int {dz d^2\mathbf{k} \over z(1-z) 16 \pi^3} \Psi_{\lambda \bar \lambda}(z,\mathbf{k})
|\ket{c_{i \lambda}(z P_+ ,\mathbf{p}_c)
\bar c^j_{\bar \lambda}((1-z)P_+,\mathbf{p}_{\bar c})} + \dots
\end{eqnarray}

\begin{figure}[h]
\centering
\includegraphics[width=0.48\linewidth]{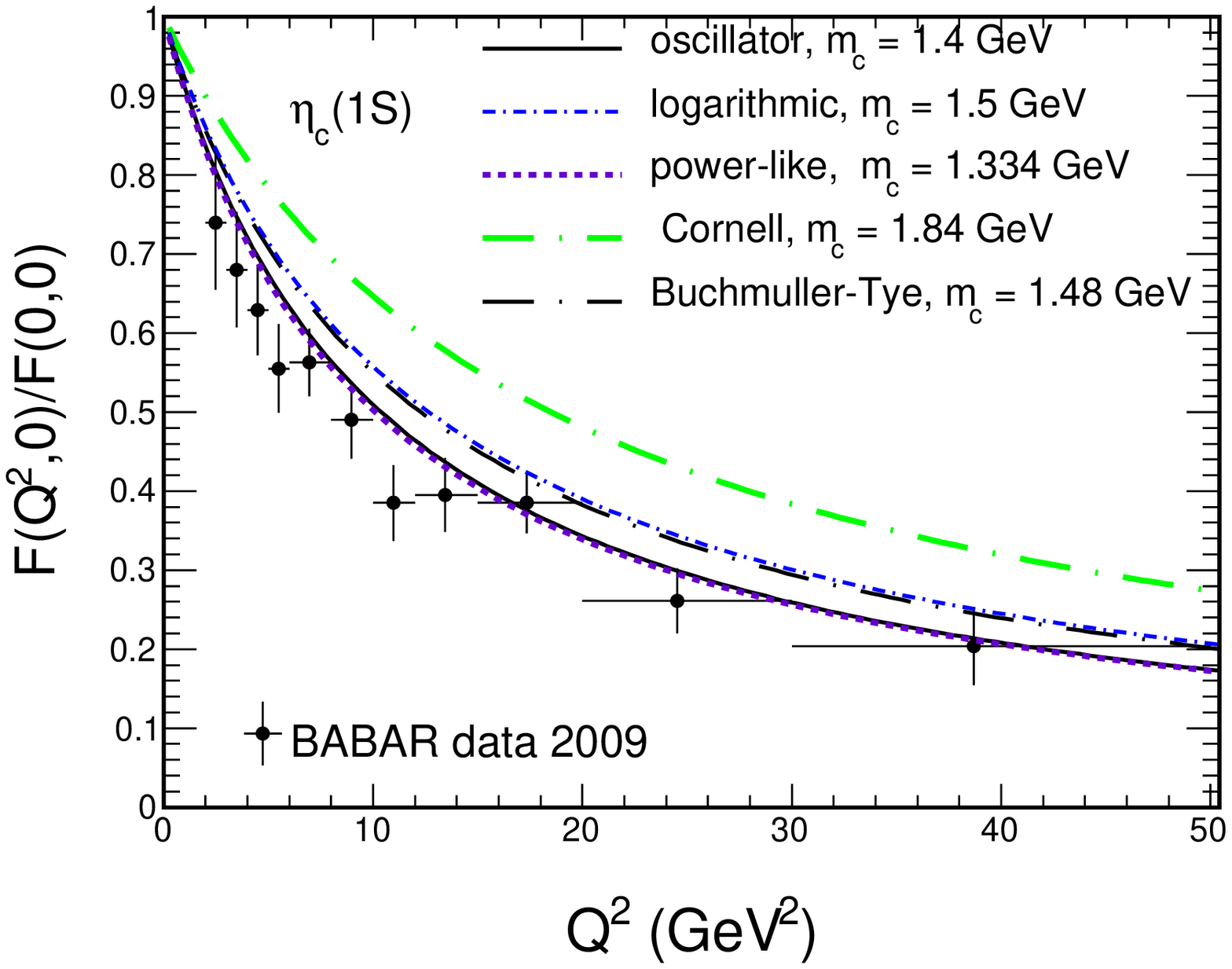}
\includegraphics[width=0.48\linewidth]{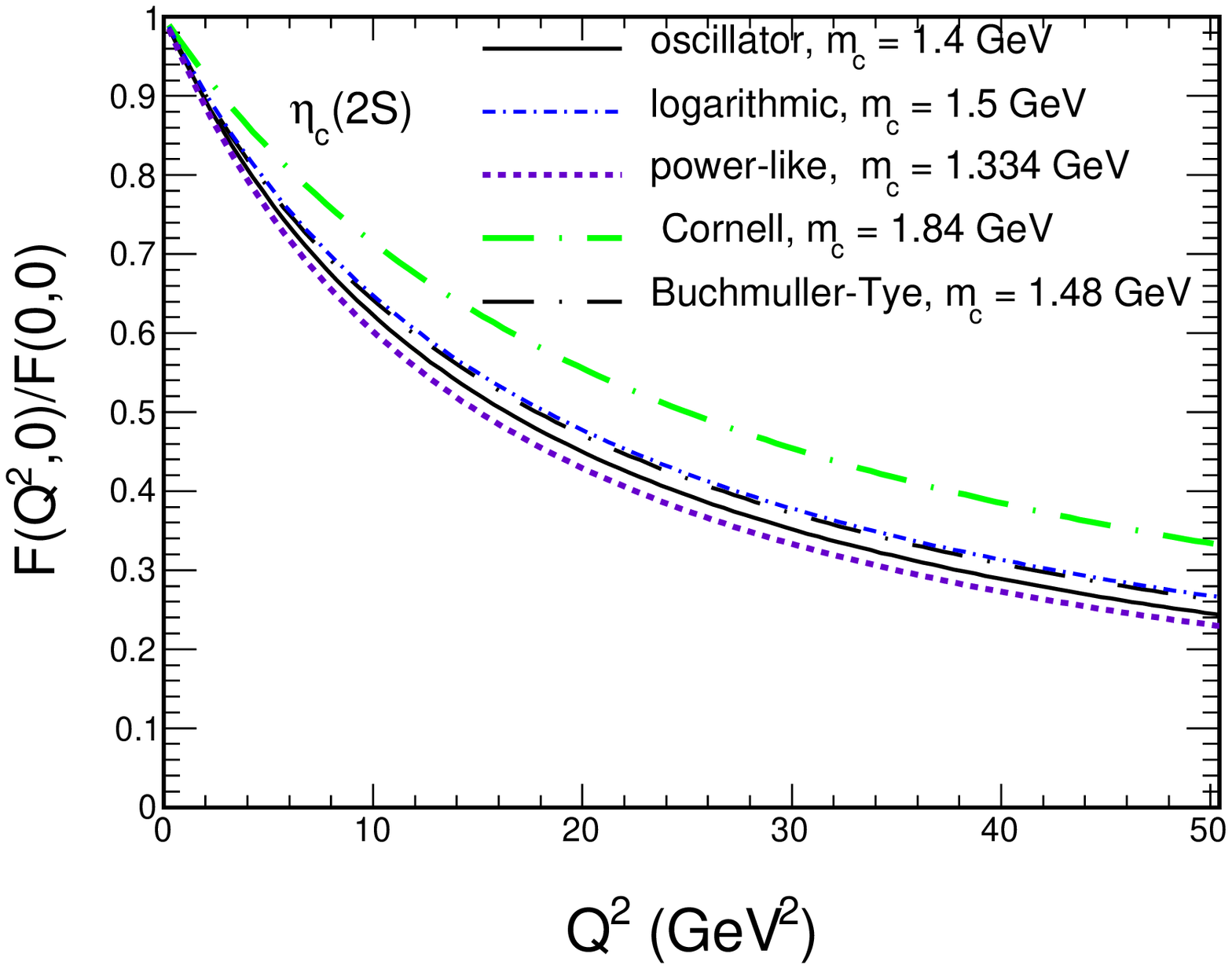}\\
\caption{Normalized transition form factor $F(Q^2,0)/F(0,0)$ as a function of photon virtuality~$Q^2$. The BaBar data are shown for comparison \cite{BaBar}.
}
\label{fig:F_Q2}
\end{figure}

We vary our numerical results by analysing several models of interaction
potential from the literature. 
In Fig. \ref{fig:F_Q2} we present normalized form factor
$F(Q^2,0)/F(0,0)$ for one real and one virtual photon compared to BaBar
data \cite{BaBar}.
The description of the experimental data appears to be related not only to applied potential model, but also to lower value of c - quark mass $m_c$.

\section{Production of $\eta_c$(1S,2S) in pp collisions}

We applied \cite{Babiarz_etac} LF form-factor in $pp$ collision by adjusting quantum numbers to gluon-gluon process. 
In the $k_T$-factorization approach, gluons are off-shell,
$q_i^2 = - \bq_i^2$ and their four momenta:

\begin{eqnarray}
q_1 = (q_{1+},0,\bq_1) \,,\, q_2 = (0,q_{2-},
 \bq_2) \, ,\, q_{1+} = x_1 \sqrt{s \over 2} \,,\, q_{2-} = x_2  \sqrt{s \over 2} \, .
\end{eqnarray}

We can write the cross section for inclusive $\eta_c (1S)$ or $\eta_c (2S)$ production in the form:

\begin{eqnarray}
d\sigma = \int {dx_1 \over x_1} \int {d^2 \bq_1 \over \pi \bq_1^2} 
{\cal{F}}(x_1,\bq_1^2,\mu_F^2)\int {dx_2 \over x_2} 
\int {d^2 \bq_2 \over \pi \bq_2^2}  {\cal{F}}(x_2,\bq_2^2,\mu_F^2) {1 \over 2 x_1 x_2 s} \overline{|{\cal{M}}|^2} \, d\Phi(2 \to 1)\, ,
\end{eqnarray}

where the phase space element: $d\Phi(2 \to 1)  = (2 \pi)^4 \delta^{(4)}(q_1 + q_2 - p) \, {d^4 p \over (2 \pi)^3} \delta(p^2 - M_{\eta_c}^2)$.

\begin{eqnarray}
{\cal{M}}^{ab} = {q_{1 \perp}^\mu q_{2\perp}^\nu \over |\bq_1| |\bq_2|}{\cal{M}}^{ab}_{\mu \nu}  = {q_{1+} q_{2-} \over |\bq_1| |\bq_2|} n^{+\mu} n^{-\nu} {\cal{M}}^{ab}_{\mu \nu} = {x_1 x_2 s \over 2 |\bq_1| |\bq_2| } n^{+\mu} n^{-\nu} {\cal{M}}^{ab}_{\mu \nu}  \, .
\end{eqnarray}

Therefore, the matrix element reads:
$n^{+\mu} n^{-\mu} {\cal{M}}^{ab}_{\mu \nu} = 4 \pi \alpha_s  (-i) [\bq_1,\bq_2]
\Tr[t^a t^b]/\sqrt{N_c} \, I(\bq_1^2,\bq_2^2)$
and averaging over colors, we obtain the final result:

\begin{eqnarray}
{d \sigma \over dy d^2\bp} = \int {d^2 \bq_1 \over \pi \bq_1^4} 
{\cal{F}}(x_1,\bq_1^2) \int {d^2 \bq_2 \over \pi \bq_2^4}  {\cal{F}}(x_2,\bq_2^2) \, \delta^{(2)} (\bq_1 + \bq_2 - \bp ) \\ 
\times {\pi^3 \alpha_s^2 \over N_c (N_c^2-1)} ||\bq_1| |\bq_2| 
\sin(\phi_1 - \phi_2) \, I(\bq_1^2,\bq_2^2)|^2 . \nonumber
\end{eqnarray}

The $I(\bq_1^2,\bq_2^2)$ above is related to the form factor for $\gamma^* \gamma^* \to \eta_c$: $F(Q_1^2,Q_2^2) = e_c^2 \sqrt{N_c} \, I(\bq_1^2,\bq_2^2) \, ,$

and going to NRQCD limit the transition form factor takes the form:

\begin{eqnarray}
F_{\rm NRQCD}(Q_1^2, Q_2^2) = {4 e_c^2 \sqrt{N_c} \over \sqrt{\pi M_{\eta_c}}} { 1 \over M_{\eta_c}^2 + Q_1^2 + Q_2^2} \, R(0) \, ,
\end{eqnarray} 
where $R(0)$ is radial wave function of the potential-model at the spatial origin.

The normalization of the cross section crucially depends on the value
of the form factor at on-shell point, thus we extract $F(0,0)$ from
experimental value of the radiative decay width (see Table \ref{table:width_gamma}) in leading order:
\begin{equation}
\label{eq:L0_photon}
    \Gamma_{\rm{LO}}(\eta_c \to \gamma \gamma) = {\pi \over 4} \alpha^2_{\rm em}
    M^3_{\eta_c} |F(0,0)|^2 \, .
\end{equation}
and the expression for the width at Next Leading Order, 
according to Ref.~\cite{Lansberg:2006dw}, is:

\begin{eqnarray}
\Gamma_{\rm{NLO}}(\eta_c \to \gamma \gamma) &=& \Gamma_{\rm{LO}}(\eta_c \to \gamma \gamma) \, \Big( 1 - {20 - \pi^2 \over 3} {\alpha_s \over \pi} \Big)\,.
\label{eq:NLO}
\end{eqnarray}

\begin{table}[h!]
    \centering
    \begin{tabular}{c|c|c|c}
    \hline
    \hline
      &Experimental values& Derived from LO &Derived from NLO \\
      &  $\Gamma_{\gamma \gamma}$(keV)\cite{PDG_2018} &$|F(0,0)| [GeV^{-1}]$  & $|F(0,0)|_{\gamma \gamma}[GeV^{-1}]$\\ 
                       \hline
    $\eta_{c}(1S)$  & 5.0 $\pm$0.4  & 0.067$\pm$0.003 & 0.079$\pm$0.003 \\
    $\eta_{c}(2S)$  & 1.9 $\pm$1.3 $\cdot$10$^{-4}\cdot \Gamma_{\eta_{c}(2S)}$ & 0.033$\pm$0.012 & 0.038$\pm$0.014\\
    \hline
    \hline
    \end{tabular}
    \caption{Radiative decay widths as well as $|F(0,0)|$ obtained from $\Gamma_{\gamma \gamma}$ using leading order and next-to-leading order approximation.}
    \label{table:width_gamma}
\end{table}

\begin{figure}[h!]
    \centering
    \includegraphics[width=0.3\textwidth]{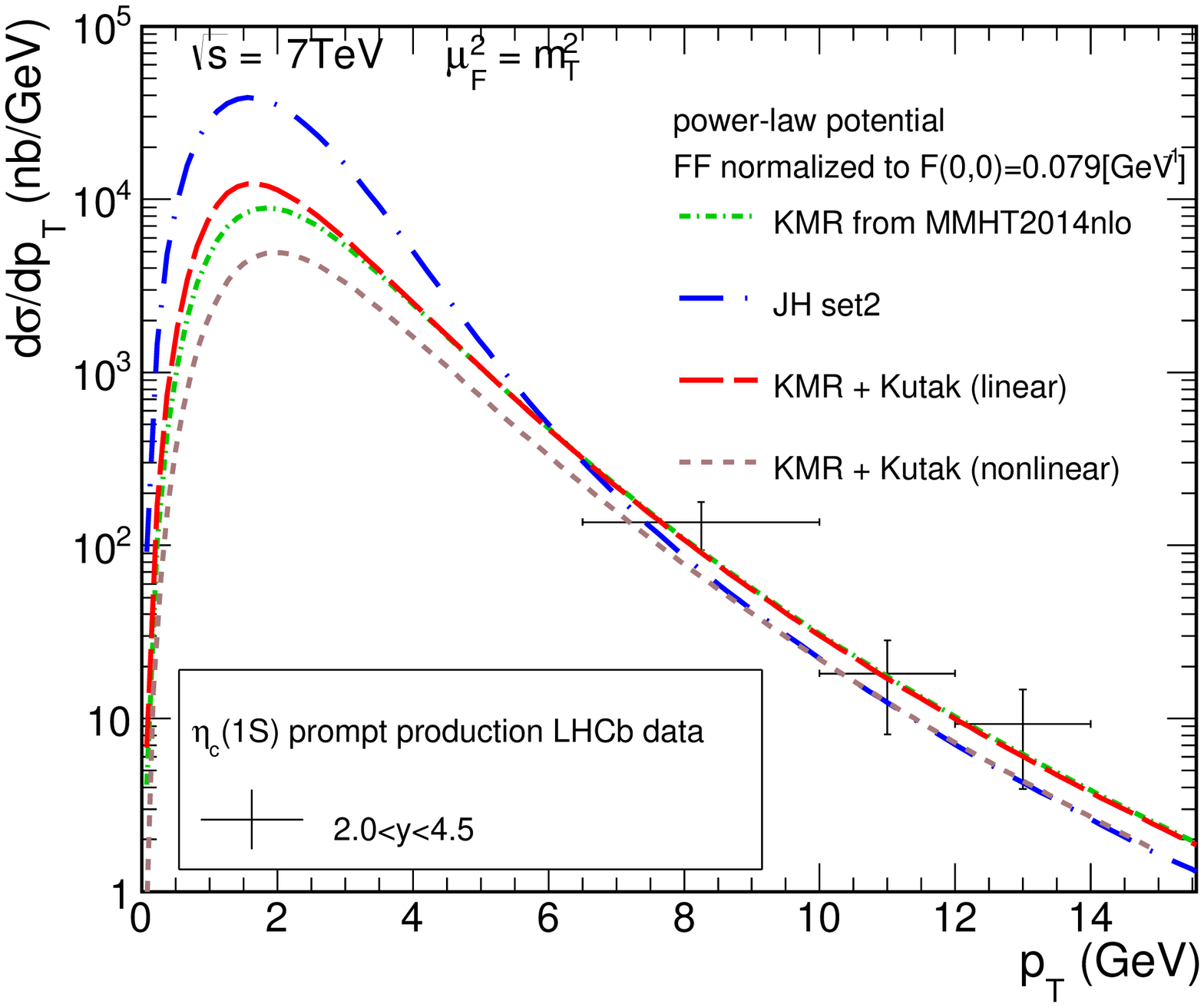}
    \includegraphics[width=0.3\textwidth]{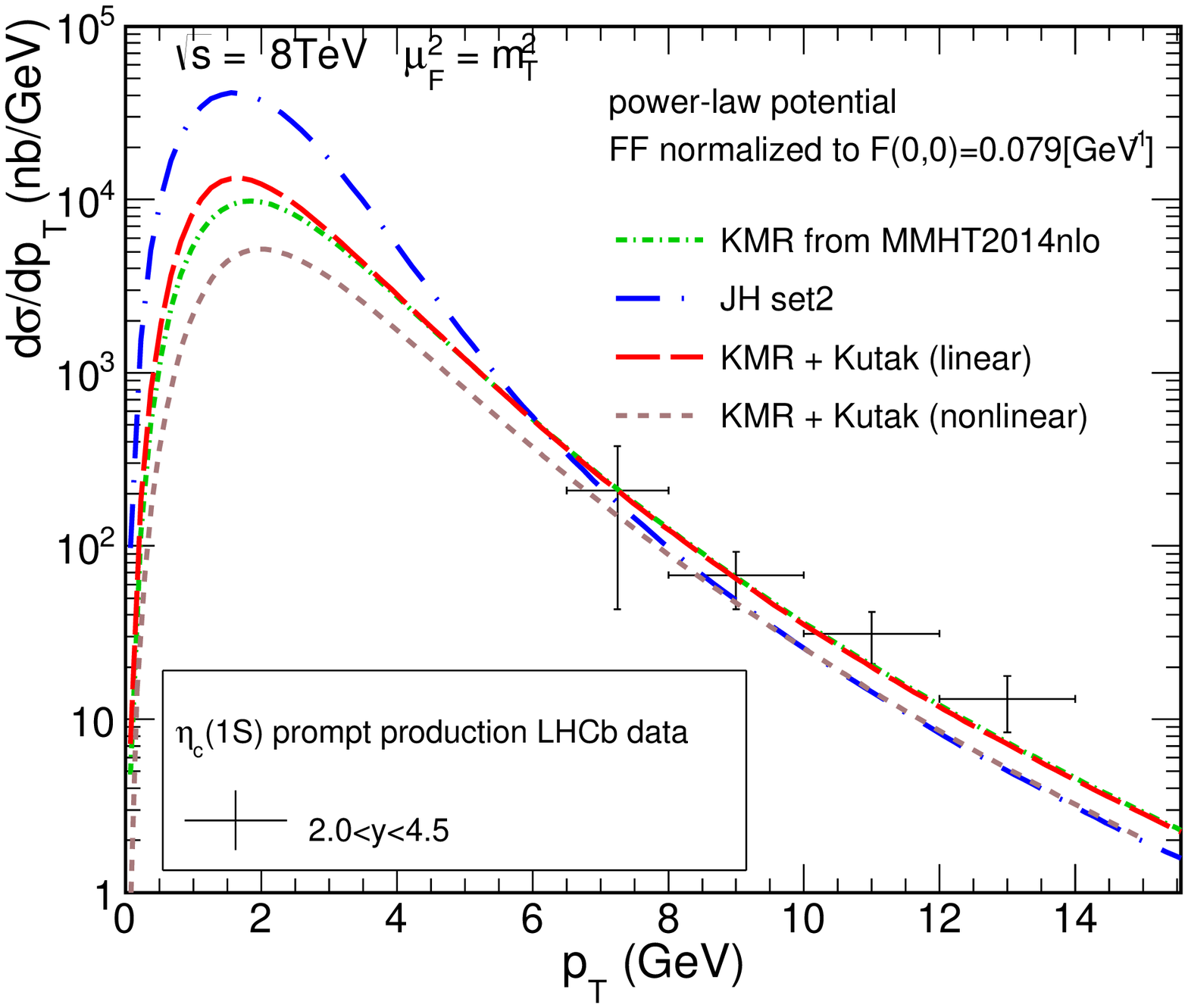}
    \includegraphics[width=0.3\textwidth]{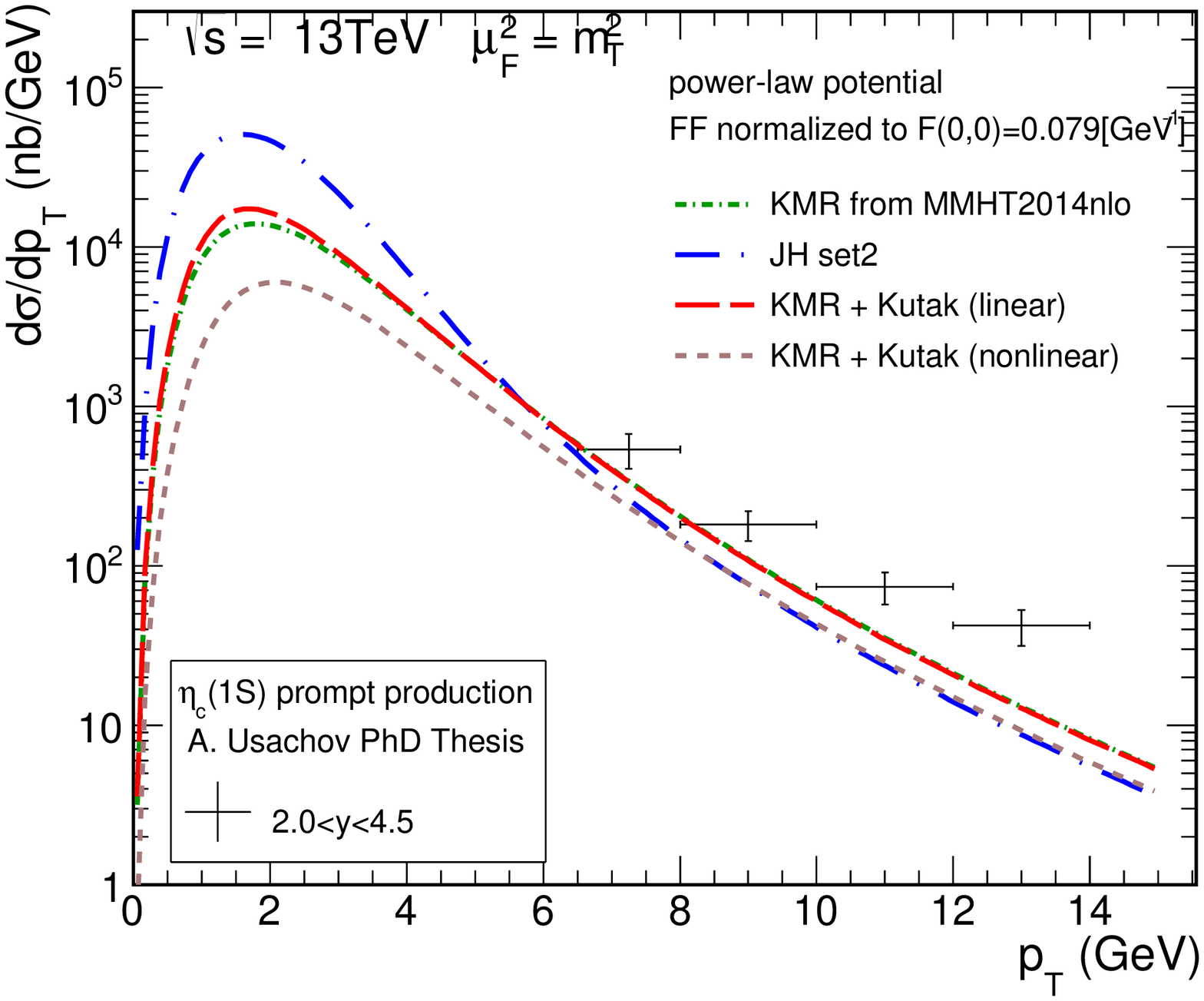}
\caption{Differential cross section as a function of transverse momentum for prompt $\eta_c(1S)$
production compared with the LHCb data for $\sqrt{s}= 7,8\, \rm{TeV}$
\cite{LHCb_2015} \cite{Usachov_2019}.}
\label{fig:dsig_dpt_etac1S}
\end{figure}

\begin{figure}[h!]
    \centering
    \includegraphics[width=0.3\textwidth]{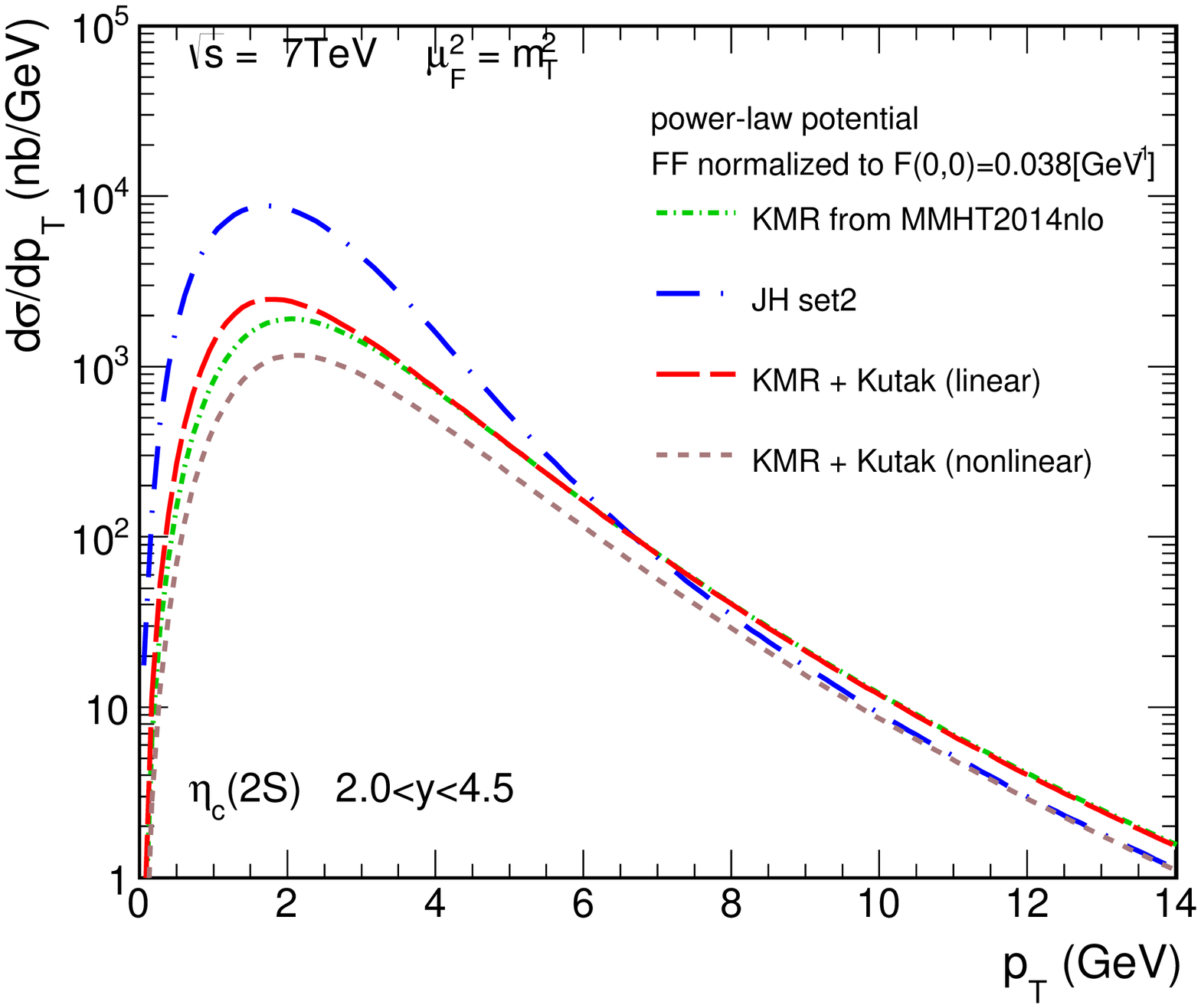}
    \includegraphics[width=0.3\textwidth]{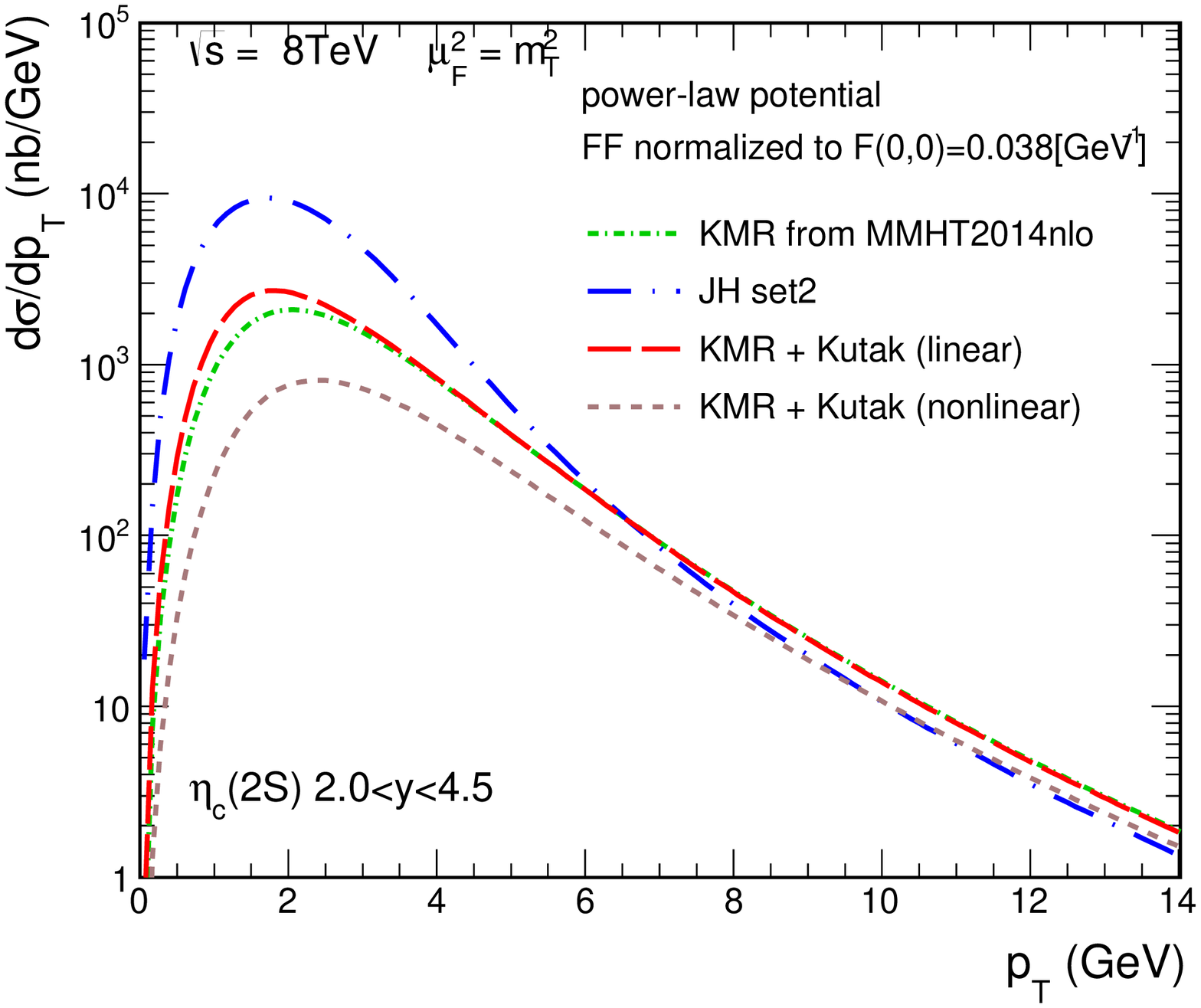}
    \includegraphics[width=0.3\textwidth]{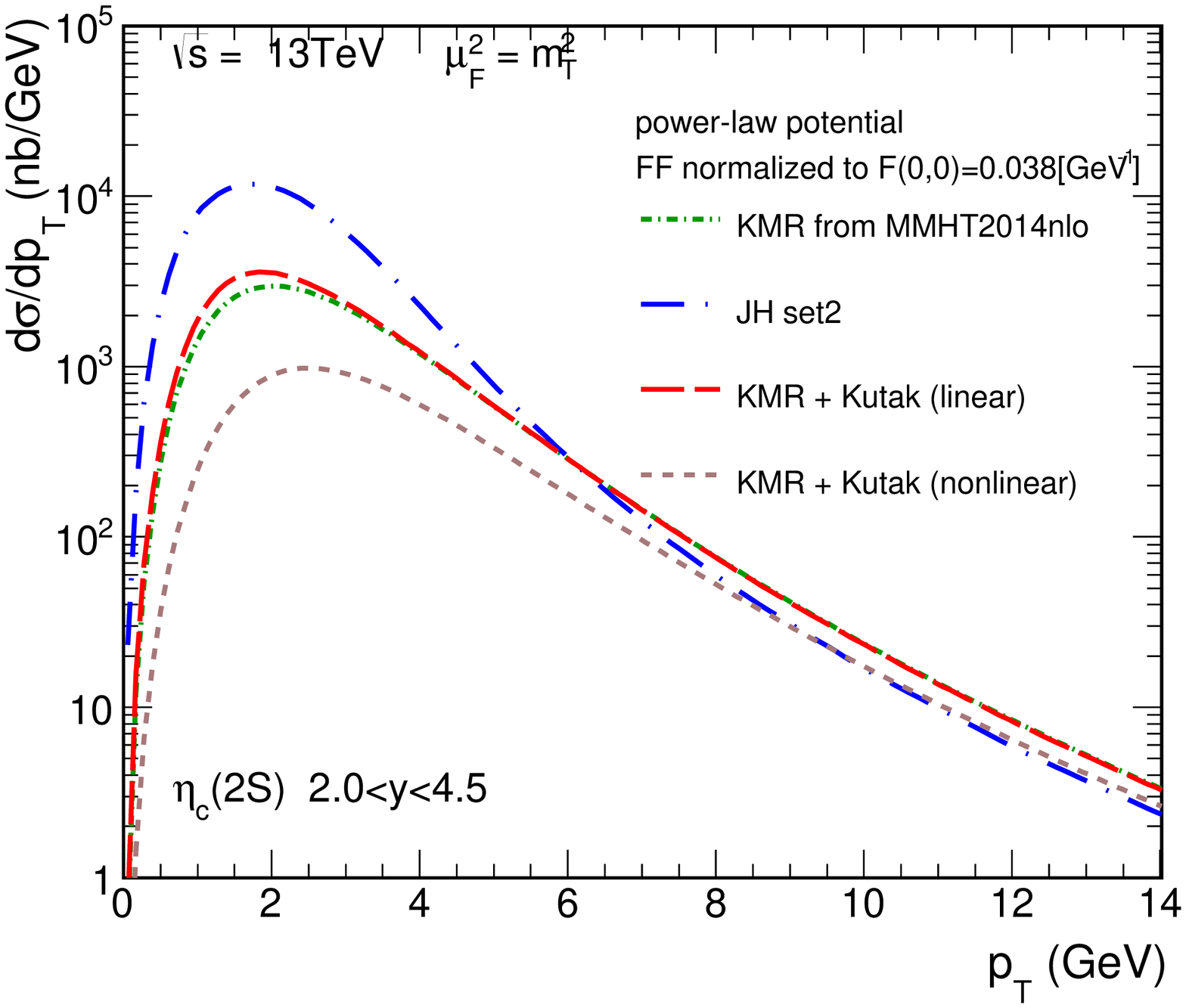}
    \caption{ Differential cross section as a function of transverse
      momentum for prompt $\eta_{c}(2S)$
      production for $\sqrt{s} = 7, 8, 13 \, \rm{TeV}$.}
          \label{fig:dsig_dpt_etac2S}
\end{figure}

\begin{figure}[h!]
    \centering
    \includegraphics[width=0.3\textwidth]{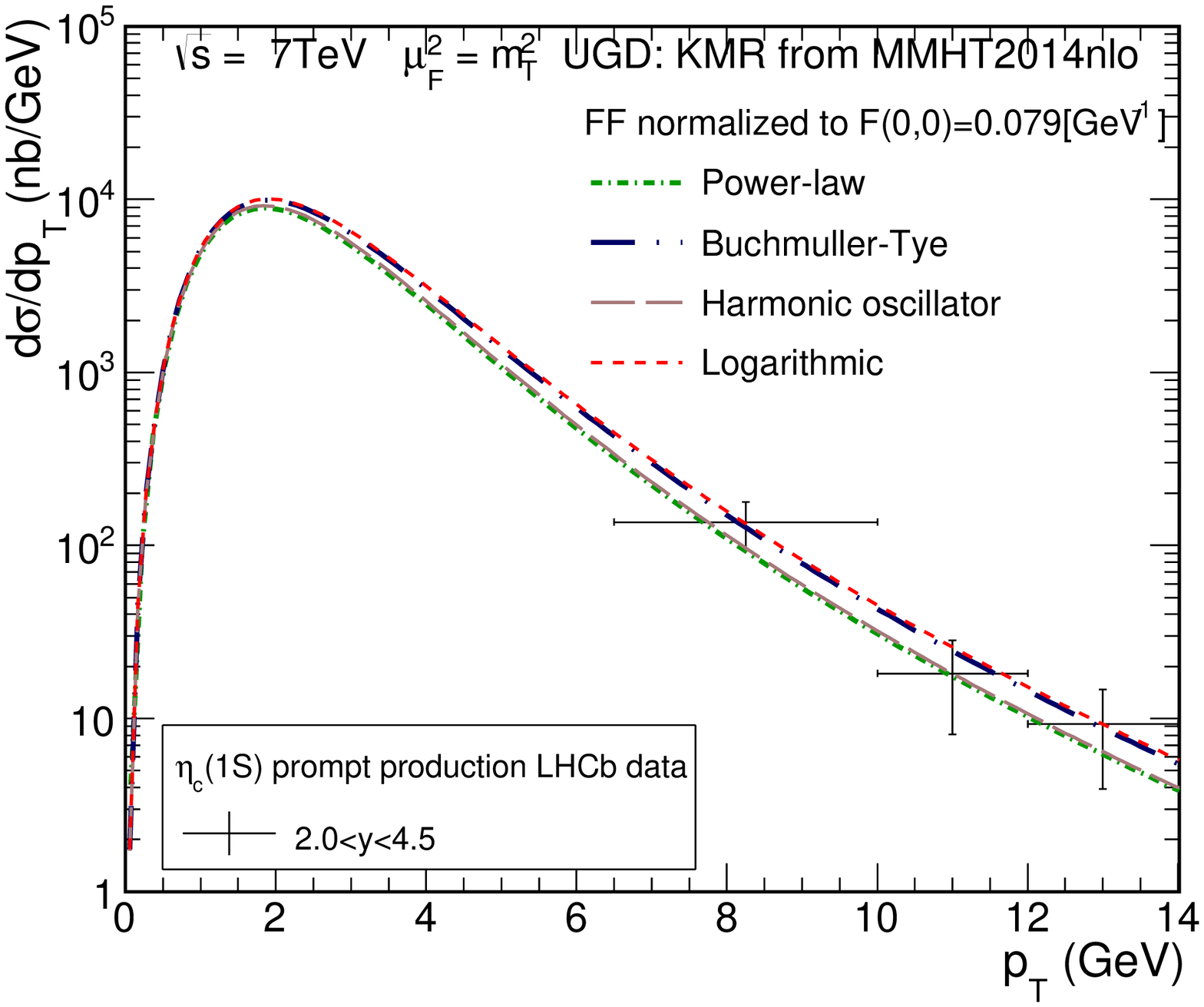}
    \includegraphics[width=0.3\textwidth]{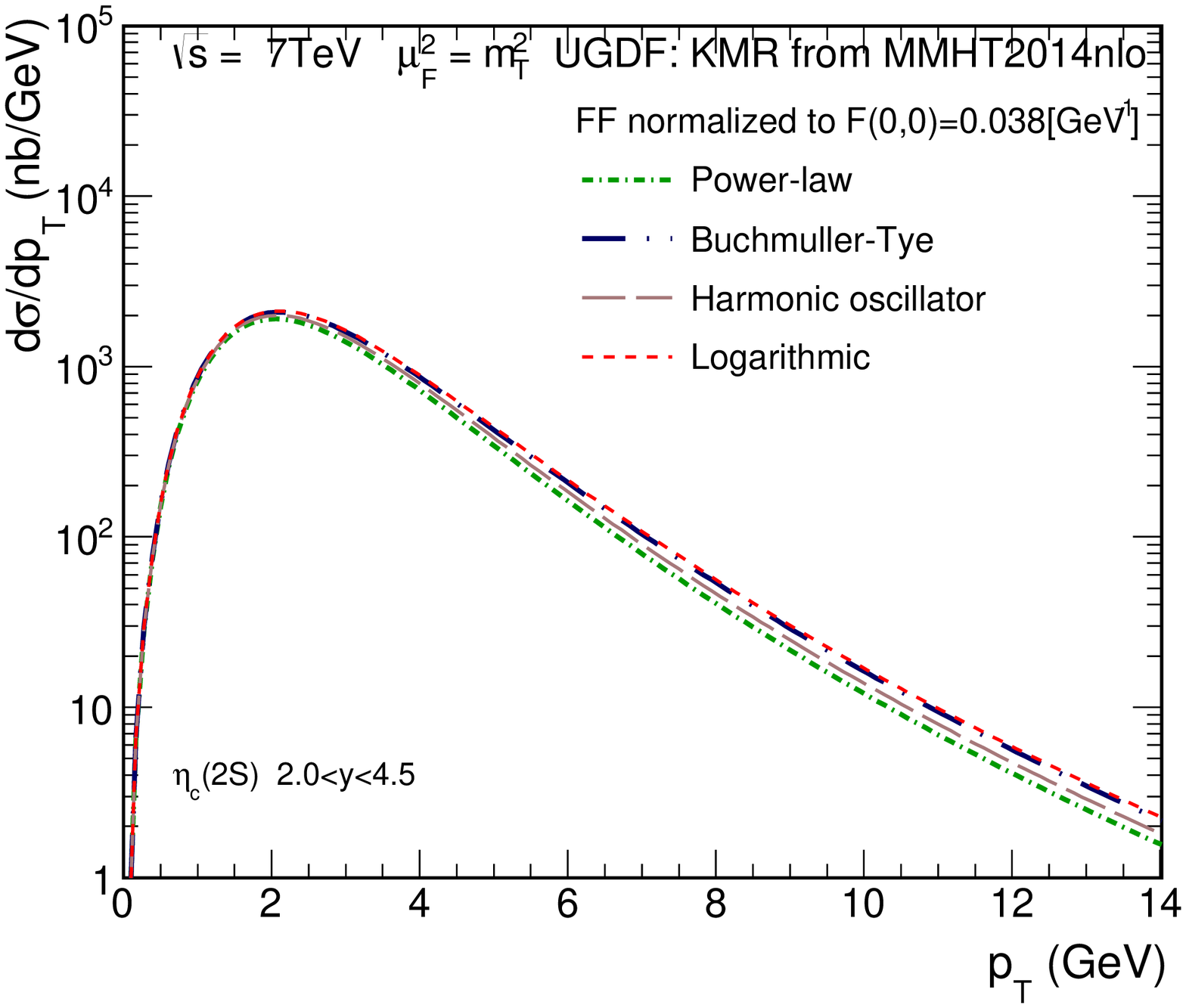}
    \includegraphics[width=0.3\textwidth]{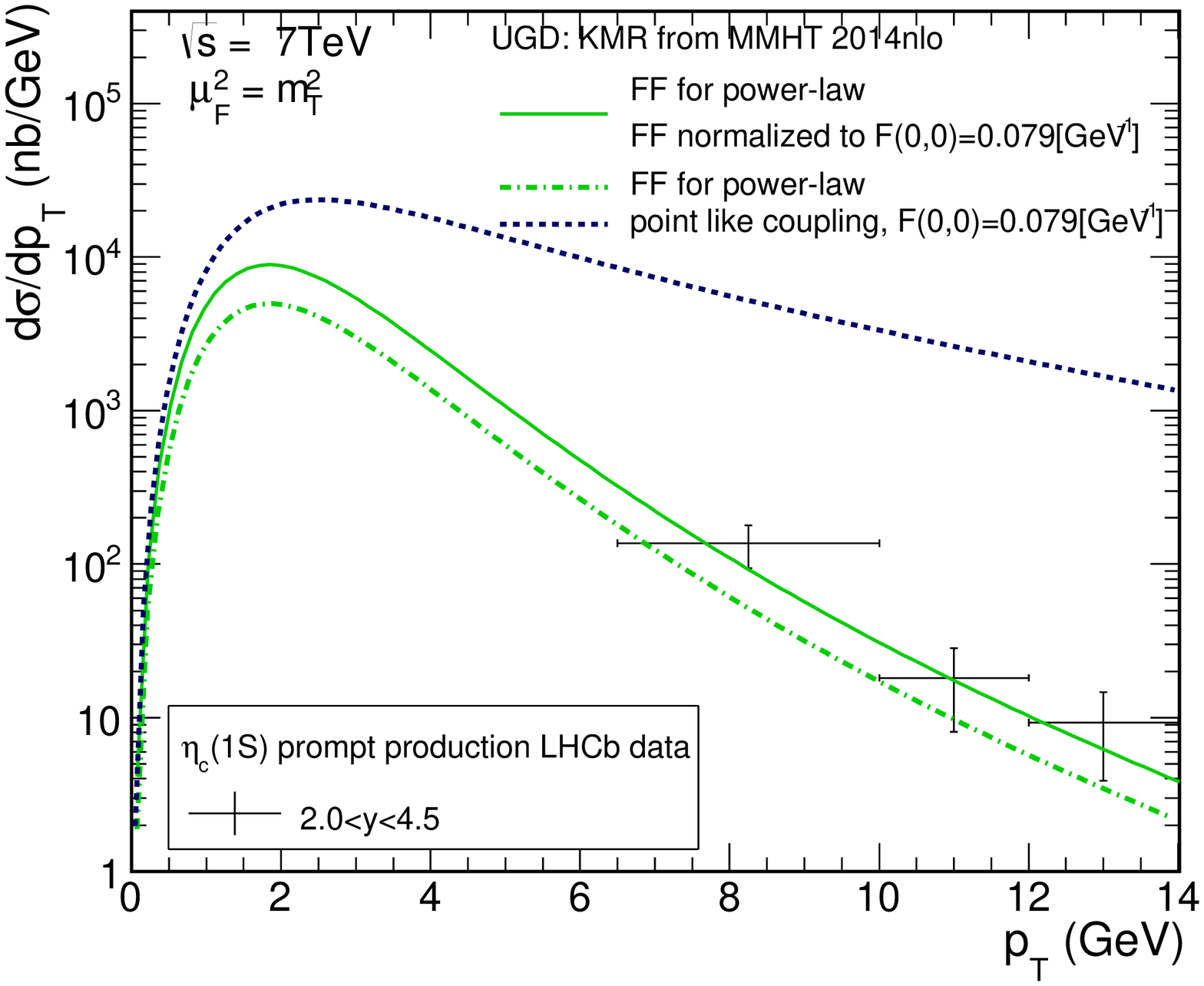}
    \caption{
    Distribution in transverse momentum of the $\eta_c(1S)$ (left panel)
    and the $\eta_c (2S)$ (middle panel) for form factor calculated,
    from different potential models, with the same normalization at the
    on shell point. On the right panel comparison of the results with
    different normalization of the form factor for the Power-law potential.}
    \label{fig:dsig_dpt_ff_exp}
\end{figure}

In Figs.~\ref{fig:dsig_dpt_etac1S} and \ref{fig:dsig_dpt_etac2S} we present differential cross section as a function 
of transverse momentum for prompt $\eta_{c}(1S)$ and $\eta_{c}(2S)$ production compared with the LHCb data 
for $\sqrt{s} =7, 8 \, \rm{TeV}$ \cite{LHCb_2015} and preliminary experimental data for $\sqrt{s}$ = 13 TeV\cite{Usachov_2019}
for the interval in rapidity $2.0 < y < 4.5$.
In the numerical calculation we used several unintegrated gluon distributions and we applied form factor calculated from 
the power-law potential as explained in the section~\ref{sect:Intro}. 
In Fig.~\ref{fig:dsig_dpt_ff_exp} we show the transverse momentum distribution
of the $\eta_c(1S)$ (left panel) and $\eta_c(2S)$ (middle panel) with form factor
obtained from wave functions for different potential models and the same normalization at the on shell point of the form factor. 

In the right panel of Fig.~\ref{fig:dsig_dpt_ff_exp}
we present the results with different normalization of the form factor,
normalized to experimental value, exact value from light front wave
functions and point like form factor. We wish to point out that it is
important to take into account gluon virtualities in the $\eta_c$ prompt
hadroproduction. This kind processes are also a good probe of Unintegrated Gluon Distributions.

\section*{Acknowledgement}
This study was partially supported by the Polish National Science Center under grant No. 2018/31/B/ST2/03537.


\end{document}